# A novel method for depicting academic disciplines through Google Scholar Citations: The case of Bibliometrics

**Alberto Martín-Martín[1], Enrique Orduna-Malea[2], and Emilio Delgado López-Cózar[1*]**

[1] Facultad de Comunicación y Documentación. *Universidad de Granada, 18071 Granada, Spain*
[2] *Universitat Politècnica de València. Camino de Vera s/n, Valencia 46022, Spain*

*edelgado@ugr.es

**Abstract:** This article describes a procedure to generate a snapshot of the structure of a specific scientific community and their outputs based on the information available in Google Scholar Citations (GSC). We call this method MADAP (Multifaceted Analysis of Disciplines through Academic Profiles). The international community of researchers working in Bibliometrics, Scientometrics, Informetrics, Webometrics, and Altmetrics was selected as a case study. The records of the top 1,000 most cited documents by these authors according to GSC were manually processed to fill any missing information and deduplicate fields like the journal titles and book publishers. The results suggest that it is feasible to use GSC and the MADAP method to produce an accurate depiction of the community of researchers working in Bibliometrics (both specialists and occasional researchers) and their publication habits (main publication venues such as journals and book publishers). Additionally, the wide document coverage of Google Scholar (specially books and book chapters) enables more comprehensive analyses of the documents published in a specific discipline than were previously possible with other citation indexes, finally shedding light on what until now had been a blind spot in most citation analyses.

## 1. Introduction

Science, in order to be properly investigated, grasped, and taught, has usually been organized in various areas of knowledge. Over time, each of these areas has been further divided into fields, subfields, disciplines, and specialties, as a result of the ever faster growth of knowledge and the parallel increase in the number of people who form the scientific communities within each of these areas. This process of scientific budding resembles the life cycle of a living being (birth, growth, reproduction, and death), and is subject to an endless metamorphosis.

Each of these units in which scientific knowledge is structured has its own idiosyncrasies and epistemological properties (its object, its principles, and its methods) that endow them with a characteristic identity as well as boundaries that demarcate their cognitive territory. However, the inner and outer boundaries are not always clearly defined due to overlaps between disciplines, gaps, and loops, sometimes quite vague and difficult to trace.

The different areas of knowledge are populated by communities of scientists and professionals, each group using their own tools, methodologies and techniques. These are social groups that share – with more or less consensus – professional practices, forms of work organization, living conditions, social expectations, principles, values, and beliefs.

Whitley (1984) dissected the process by which academic communities – and their disciplines and specialties – become socially and cognitively institutionalized: how they create organizations that allow them to associate in order to defend their interests; how they erect spaces for the exchange of ideas and social development (conferences, seminars, forums, etc.); how they institute professional (newsletters, discussion lists) or scientific means (journals) of communication; how they obtain academic standing by teaching the subject at the university (courses in graduate and postgraduate programs, including Master and PhD degrees); how they create groups, departments, laboratories, and companies dedicated to advance research; how they define research agendas where not only research problems but also ways to solve them are addressed; or how to create a common language to establish ideas and principles. Not to mention that the process of social and cognitive institutionalization of disciplines is directly influenced by the geographic location and the different levels of economic and cultural development of the countries where researchers are based.

As formulated by Becher and Trowler (2001), there is a close relationship between the disciplines (territories of knowledge) and people who advance them (scientific tribes); between the epistemic properties of the forms of scientific knowledge and the social aspects of academic communities. This is why any analysis of a discipline cannot ignore the cognitive (disciplines themselves) and social (community) areas. A discipline is what is performed by those who cultivate it.

Being aware of the scope of a discipline will not only help characterize and determine its perspective and scientific nature, but it will also indirectly delineate its internal structure, its coherence, its contours, and its location in the overall picture of the Sciences. This will enable an understanding of what the research is and has been about in a particular discipline, and how it may evolve in the future.

Although there is no unanimity yet about what the most appropriate methods to describe disciplines are, this work intends first to depict one scientific discipline and those who practice it (through a multifaceted approach based on the intellectual production generated by its academic community), and second, to carry out this procedure using both semi-supervised (Google Scholar Citations) and unsupervised environments (Google Scholar).

Therefore, the main goal of this work is to investigate the suitability of Google Scholar (GS) and Google Scholar Citations (GSC) to provide a comprehensive and multifaceted picture of the structure of an entire scientific specialty through the main agents that are part of it (scientists, professionals, the documents they produce, and the venues where these documents are published).

While classic citation indexes (Scopus and Web of Science) have been traditionally used to analyse scientific disciplines, their particular coverage and principles (controlled sets of journals that represent the elite, based on a Bradford-like core) have probably constrained the pictures that could be obtained. These databases provide a better coverage in areas like Science, Medicine and Technology, but they lack many relevant sources in areas like the Social Sciences and Humanities. Academic search engines like GS practice a radically different approach when it comes to selecting sources to cover and index, and therefore it might be useful to explore the wider view of academic outputs that they provide (Martín-Martín et al. 2016). To the best of our knowledge, there have not yet been any attempts to comprehensively analyse an entire discipline using GS and GSC.

Both GS and GSC present a series of well-known shortcomings and restrictions that hinder the use of these platforms for bibliometric analyses (Jacsó 2005; 2008; 2012; Meho and Yang 2007; Aguillo 2012; Prins 2016). Therefore, the development of a method that enables the use of these platforms for bibliometric purposes would facilitate studies that are not limited by the document coverage biases of other citation indexes.

In this line, this study intends to answer the following questions:

> RQ1: Can GSC and GS be used to generate a representation of the community of authors that work in any given academic discipline, and their outputs?
>
> RQ2: Is it possible to apply a multi-faceted approach to analyse a discipline with the data available in GS and GSC?

A positive answer to these questions would mean that it is possible to carry out bibliometric analyses of disciplines using Google Scholar Citations, a source of data that is free to access and semi-automatically updated. The data from this source could at the very least complement the data available in other subscription-based citation indexes.

In order to answer this research question, this work takes as a case study a very specific scientific and professional community (Bibliometrics) along with its close-related areas (Scientometrics, Informetrics, Webometrics, and Altmetrics). The reason behind the selection of this discipline is that the authors are familiar with this field. This expertise is considered necessary in order to assess the results of the analyses and be able to detect the potential shortcomings of the method.

## 2. Research background

### *The object of study. Bibliometrics: A discipline with many names*

There are numerous works which address the history of Bibliometrics (Broadus 1987a; Hertzel 1987; Shapiro 1992; Godin 2006; De Bellis 2009). Its denomination, object of study and scope have been addressed as well (Lawani 1981; Bonitz 1982; Peritz 1984; Broadus 1987b; Brookes 1988; 1990; Sengupta 1992; Glänzel and Schoepflin 1994; Braun 1994; Gorbea 1994; Hood and Wilson 2001; Cronin 2001; Thelwall 2008; Lariviere 2012). There are also several literature reviews about this subject (Narin and Moll 1977; White and McCain 1989; Van Raan 1997; Wilson 1999; Borgman and Furner 2002).

Bibliometrics can be synthetically defined as the discipline responsible for measuring communication and, more specifically, as the specialty responsible for quantitatively studying the production, distribution, dissemination and consumption of information conveyed in any type of document (book, journal, conference, patent, or website) and across all spheres of activity, but with special attention to scientific information. This discipline has various peculiar features:

a) It is a very young discipline, and its epistemic foundations are still not fully defined.
b) It is a discipline best defined by its methods than by the thematic areas that it covers.
c) It has a strong interdisciplinary nature, which arises from the incorporation of methods and techniques developed in other fields, and by its application to the study of any subject area.

It is probably because of these reasons that this discipline is known by many different names. However, this fact does not mean that the subject of study or the borders of the discipline are not clearly defined. Rather, it is a sign of the coexistence of different traditions that have shaped the development of the discipline.

Bibliometrics is the original and most widely-used term to refer to it. It stems from the bibliographic tradition represented by Paul Otlet with his proposal for a "bibliometrie", a Science for measuring all the dimensions of books and other documents (Otlet 1934), and from the library tradition concerned since ancient times with measuring the growth of knowledge and the usage of its holdings (Ranganathan 1969).

Scientometrics is oriented towards the quantitative analysis of scientific and technical literature. It comes from the tradition of the science of science (space of confluence of Sociology, History, and Philosophy of science), to which science policy is also linked. It was crucial for this scientometric orientation the creation of the citation indexes (Garfield 1970).

Informetrics is focused on the discovery of mathematical models that explain the properties of information (Egghe and Rousseau 1990; Tague-Sutcliffe 1992; Bar-Ilan 2008). It is connected with the modern information science. Webometrics (Almind and Ingwersen 1997; Thelwall, Vaughanand Björneborn 2005; Thelwall 2009) and Altmetrics (Priem and Hemminger 2010) are the most recent denominations. They started to gain momentum as the use of the new information and communication technologies began to spread. They are being developed in the tradition of the modern Library and Information Science, a discipline increasingly dedicated to computer science and to computing itself. These new names are strongly influenced by the medium in which information is conveyed rather than by the content itself.

The terms used as well as their conceptual domains and boundaries have been already described in the literature (Björneborn and Ingwersen 2004; Milojević and Leydesdorff 2013; Stuart 2014). However, there is no consensus on the precise relation among them. By way of illustration, an analysis of the five selected terms (Bibliometrics, Scientometrics, Informetrics, Webometrics and Altmetrics) used in the titles of documents published between 1969 and 2016 and indexed in GS (Figure 1) shows a clear predominance of the term “Bibliometrics”, followed by “Scientometrics”.

**Figure 1. Frequency of the terms "Bibliometrics", "Scientometrics", "Informetrics", "Webometrics" and "Altmetrics" in the title of documents indexed in Google Scholar (1969-2016)**

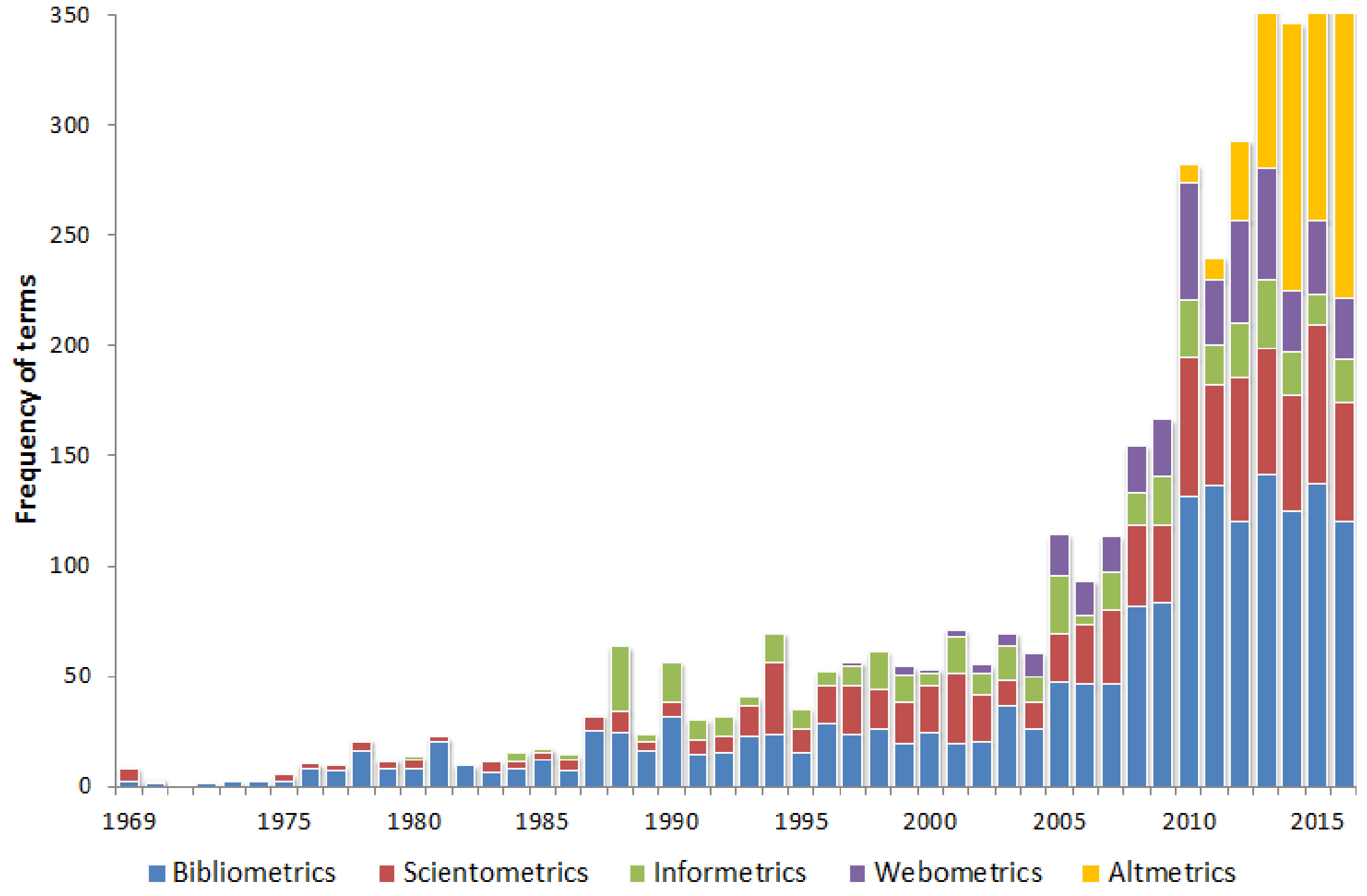

The term "Altmetrics" is being increasingly used (Figure 2) in the last three years as a result of the novelty of the new social media communication technologies. Another reason why Altmetrics is currently a hot topic in the field is the relatively unknown role that the metrics that this term encompasses can play in the quantification and evaluation of academic impact, both at the article (Lin and Fenner 2013) and author levels (Bar-Ilan et al. 2012; Orduna-Malea et al. 2016).

**Figure 2. Interest measured in search queries frequency of the terms Bibliometrics, Scientometrics and Altmetrics**

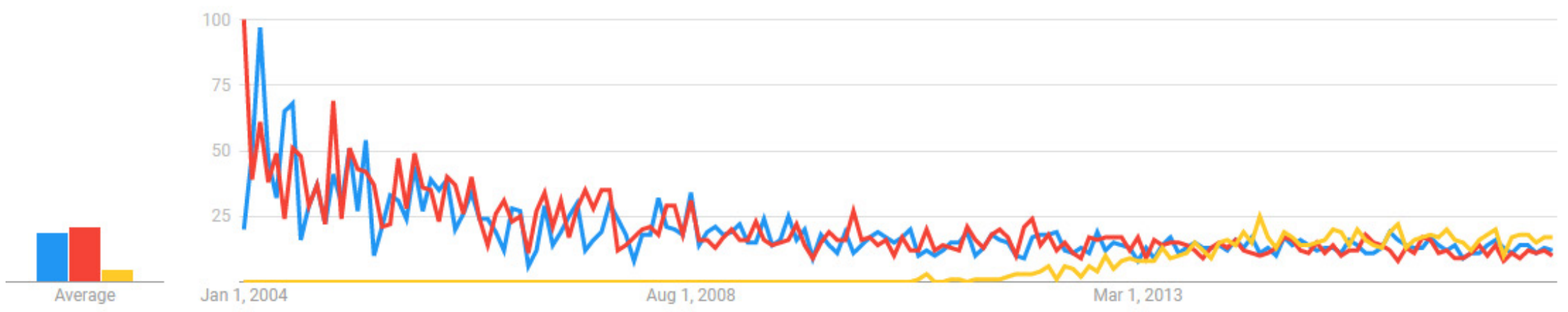

**Source: Google Trends**
**Blue: Bibliometrics; Red: Scientometrics; Yellow: Altmetrics**

***The unit of analysis. Google Scholar Citations: an unmoderated academic profile***

GSC was launched in 2011 (Jacsó 2012) and currently stands out as one of the preferred academic profiles by scholars. Kramer and Bosman (2015) released a comprehensive report about the use of academic communication tools, finding that GSC was used by 62% of the surveyed users (about 20,000), in second place just after ResearchGate (66%). The fact that GSC is linked to Google Scholar, currently the most comprehensive academic bibliographic database (Orduna-Malea et al. 2015), as well as the preferred source to start academic information discovery processes (Orduna-Malea et al. 2016), makes this service an essential professional tool for academics.

Several studies have recently used data extracted from GS for bibliometric purposes (Bornmann, Thor, Marx, and Schier 2016; Martín-Martín, Orduna-Malea, Ayllón and López-Cózar 2014; Mingers and Meyer 2017; Mingers, O'Hanley and Okunola in press). However, since the information contained in GSC is better structured than in GS, this platform has recently started to be used as a new source for bibliometric studies. Ortega and Aguillo (2012) used GSC to map the labels included in each profile to build a Science map as well as to construct country and institutional collaboration networks using co-authors lists of these profiles (Ortega and Aguillo 2013). The issue of its coverage has been addressed as well, finding not only an unbalanced subject coverage (with an important bias in favour of Computing Sciences and Engineering) but also a bias in favor of young researchers and specific institutions and countries (Ortega 2015a). Despite this, Ortega and Aguillo (2014) acknowledge that GSC has an interesting potential for research evaluation, such as a wider coverage of academic outputs and therefore a broader coverage of research impact.

Some other studies have applied GSC data to specific research environments. For example, Ortega (2015b) focused on the researchers affiliated to the Spanish National Research Council (CSIC), and Mikki et al. (2015) focused on the researchers at the University of Bergen. Nevertheless, these studies only analyse specific institutions.

Haustein et al. (2014) studied the social media presence of attendees at the 2010 STI conference celebrated in Leiden (57 researchers, who together had authored 1,136 papers). However, to the best of our knowledge there has not yet been any exhaustive study focused on one academic discipline (in this case Bibliometrics), which addresses not only author-level metrics but also documents and sources. Therefore, the main objective of this paper is to identify and describe a scientific discipline through the data available in GSC on the authors who work in said discipline.

## 3. Methods

We developed and tested a method to capture, classify and measure data from the different scientific agents of one discipline. We called this method MADAP (Multifaceted Analysis of Disciplines through Academic Profiles).

### 3.1. Author profiles search and identification

The first step was to identify all authors who have published in the areas of Bibliometrics, Scientometrics, Informetrics, Webometrics or Altmetrics, and for whom a GSC public profile could be found at the time of data collection (July 24th 2015). In order to identify the set of authors relevant to our study, an iterative snowball process was conceived, which consisted on the following search strategies.

a) Keywords

A search was conducted in four core selected journals (Scientometrics, Journal of Informetrics, Research Evaluation, and Cybermetrics) as well as the ISSI conferences (International Conference on Scientometrics and Informetrics) with the goal of extracting the most frequently used and representative words in the discipline. This process was driven by the need of capturing keywords describing the discipline. Among these terms we expected to find the most common keywords that authors use to describe their scientific interests in their GSC profiles. For this reason, we considered that these four purely bibliometric sources were sufficient for this purpose. The inclusion of other important sources which publish bibliometric studies, but also publish studies in other topics (for example, JASIST) might have introduced too much noise (keywords related to information retrieval, for example) and we think it unlikely that they would have provided any relevant terms that could not be extracted from the other journals.

To do this, the bibliographic records from all indexed articles published by these four sources were automatically retrieved using the Web of Science (n= 7143). This database was used due to its data export features, which facilitated the extraction of the documents' keyword field, a field that is not available in the metadata presented by GS. Next, all significant terms from the documents' titles and keywords (when available) were extracted. A pool of 619 terms (458 from titles and 161 from keywords) with a minimum frequency of occurrence of five in our set of documents was obtained. This vocabulary was manually processed to merge variants of the same term (for example, bibliometric and bibliometrics), delete duplicates, and exclude irrelevant terms (e.g., credit, editorial board, Nobel price, item, program, content, etc.), which were highly mentioned but useless for our purpose of representing a discipline.

After obtaining the list of terms, we checked for the existence of GS profiles in which the authors had selected one or more of these terms as their areas of interest (GSC allows authors to display up to five areas of interests). For example, the term "citation index" appeared in the title of 89 articles. However, no one had selected this term in their GS profile. Terms that no author had selected as a research interest were therefore ignored from this point on.

Lastly, the data available in all public GSC profiles that contained one or more of the selected terms as areas of interest were collected. The lack of normalization in the use of keywords sometimes forced us to search alternative keywords. These variants included misspelled words, the same keywords in other languages, etc.

b) Institutional affiliation

All the profiles associated with research centres working on Bibliometrics were also selected regardless the research interest keywords used by authors. As an example, profiles with verified e-mail domains such as <cwts.leidenuniv.nl>, <cwts.nl>, or <science-metrix.com> were selected.

c) Additional searches

Since there may have been some authors working in the discipline and who have created a public GSC profile, but who haven't added significant keywords or appropriately filled the affiliation field in their profile, we also conducted a topic search on GS (using the same previously selected terms) as well as a journal search (all the documents indexed in Google Scholar published by the core journals previously mentioned), with the aim of finding authors we might have missed with the previous two strategies.

The last two search strategies provided profiles with new keywords, some of them quite important to the discipline though they did not appear in the sample of 7143 document titles (e.g., Science and Technology Policy; 0 mentions in Titles, 72 authors including this term). These keywords were included in the final master list of disciplinary keywords. All terms that are not exclusively related to the discipline (Information Science: 61 profiles; Open Access: 41 profiles; Information literacy, 36 profiles) were excluded. The final master list of keywords consisted of 18 keywords. Table 1 displays the frequency of occurrence of these terms in the sample of documents (in Title and Keywords) and the number of authors that use that keyword in their GSC profile to describe their research interests.

**Table 1. List of Keywords describing Bibliometrics discipline**

| **Term** | **WoS source** | | **GSC Profile source** |
|---|---|---|---|
| | Title | Article Keyword | Author Keywords |
| Bibliometrics | **640** | 313 | 444 |
| Scientometrics | **372** | 127 | 382 |
| H-Index | **152** | 144 | 1 |
| Impact Factor | **135** | 149 | 1 |
| Citation Analysis | **124** | 199 | 58 |
| Informetrics | **108** | 21 | 75 |
| Research Evaluation | **62** | 104 | 74 |
| Webometrics | **38** | 49 | 68 |
| Patent Citation | **30** | 17 | 1 |
| Research Assessment | **26** | 28 | 13 |
| Citation Count | **25** | 0 | 0 |
| Research Policy | **17** | 16 | 37 |
| Science Policy | **16** | 21 | 148 |
| Altmetrics | **11** | 27 | 29 |
| Science Studies | **9** | 0 | 57 |
| Quantitative Studies of Science and Technology | **6** | 0 | 1 |
| Science Evaluation | **3** | 0 | 7 |
| Science and Technology Policy | **0** | 21* | 72 |

* Occurrences for "Science and Technology"

### 3.2. Filtering and classification of author profiles

GSC gives authors complete control over how to set their profile (personal information, institutional affiliation, research interests, as well as their scientific production). For this reason, a systematic manual revision was carried out in order to:

- Detect false positives: authors whose scientific production doesn't have anything to do with this discipline, even though they labelled themselves with one or more of the keywords associated with it.
- Classify authors in two categories:

  a) *Specialists*: authors whose scientific production substantially falls within the field of Bibliometrics.
  b) *Occasional*: authors who have sporadically published bibliometric studies, or whose field of expertise is closely related to Scientometrics (social, political, and economic studies about science), and therefore they can't be strictly considered bibliometricians.

In order to set the boundaries between the two categories (specialist and occasional authors), we decided to consider as "specialist authors" those who meet the following criterion: at least half of the documents which contribute to their h-index should fall within the limits of the field of Bibliometrics.

In order to establish the limits of the field we considered the titles of the documents as well as the venue where they were published, focusing our attention in the journals. Our Bradford-like core of journals about Bibliometrics consisted of six journals (Scientometrics, Journal of Informetrics, JASIST, Research Evaluation, Research Policy, and Cybermetrics), followed by other LIS journals which also publish numerous bibliometric studies (Journal of Information Science, Information Processing & Management, Journal of Documentation, College Research Libraries, Library Trends, Online Information Review, Revista Española de Documentación Científica, Aslib Proceedings, and El Profesional de la Información). Lastly, journals devoted to social and political studies about science (Social Studies of Science, Science and Public Policy, Minerva, Journal of Health Services Research Policy, Technological Forecasting and Social Change, Science Technology Human Values, Environmental Science Policy, and Current Science) were also searched.

811 GSC profiles were identified, out of which 48.83% (396) were classified as specialists, and the remaining 51.17% (415) as occasional authors in Bibliometrics.

### 3.3. A multi-faceted approach: units of scientific analysis

Once the set of 811 authors had been identified, we extracted the number of citations received by each of them directly from their GSC profiles (see Table 2). Additionally, we automatically extracted – by means of an ad hoc web scraper – the top 100 most cited documents for each specialist author from their GSC profile. To this set of documents (39,600), we manually added the documents we found through the additional keyword and journal queries that had been previously performed in Google Scholar (15,000 documents authored by researchers with or without a public profile in GSC).

After deleting duplicates, a set of roughly 41,000 documents remained. In the cases where various versions of the same document were found with different number of citations, the one with the highest citation count was selected. This list was sorted according to the number of citations. For each of the top 1,000 most cited documents in this list, the basic bibliographic information (especially the sources: journals and book publishers) were collected (see Tables 3, 4, and 5).

For the sake of clarity we should point out that in those cases when a book is a collective work, the number of citations is the sum of the citations to each of the chapters, in addition to the citations directed to the book as a whole.

A graphical visualization of the MADAP procedure can be found in Figure 3

**Figure 3. Description of MADAP method**

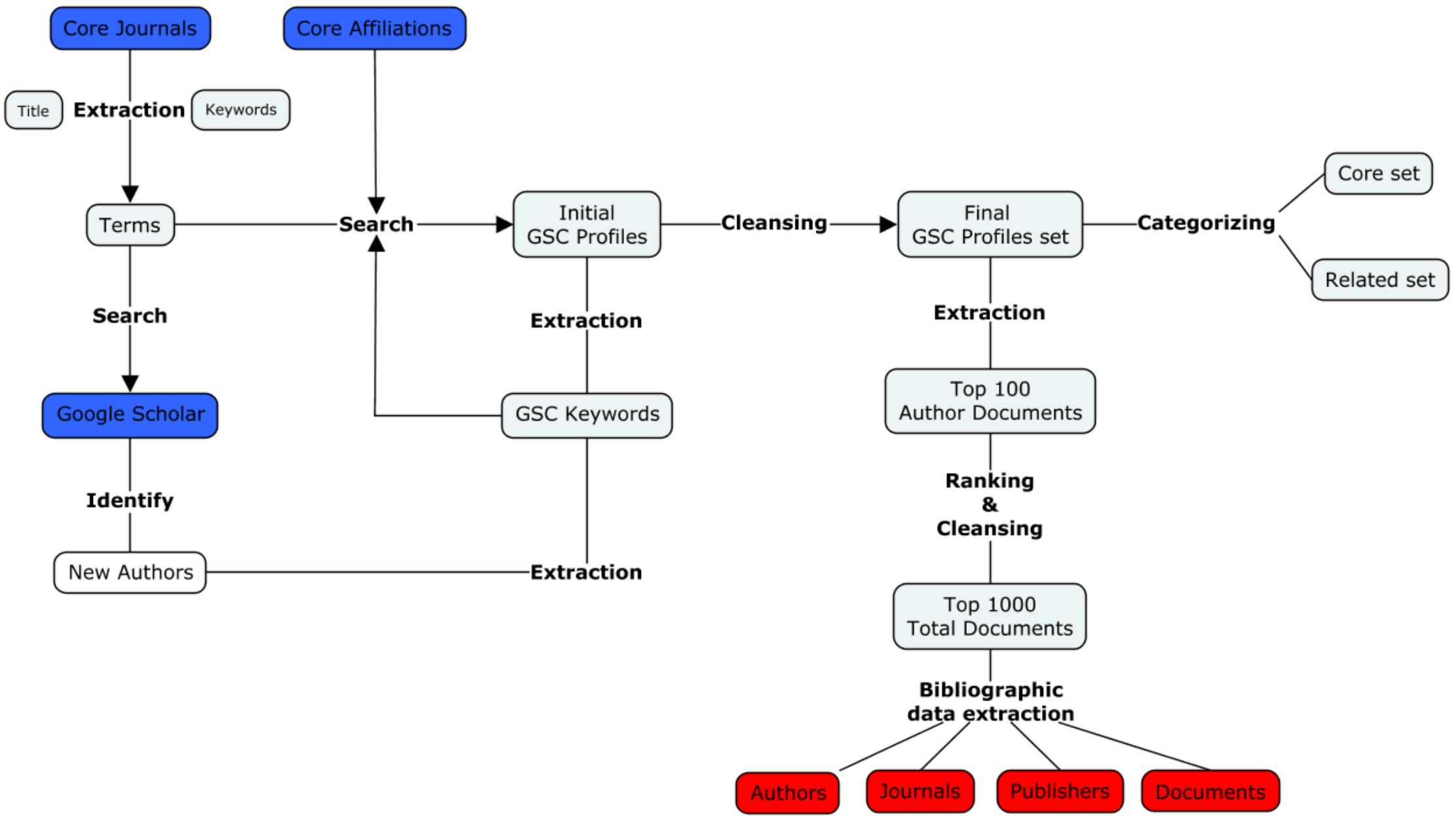

## 4. Results

### 4.1. The actors of Bibliometrics according to Google Scholar Citations, through the MADAP method

a) Authors

The list of most influential authors of the discipline is available in the Table 2.

**Table 2. Top 25 influential specialist/occasional authors in Bibliometrics according to Google Scholar Citations**

| SPECIALIST AUTHORS | CITATIONS | H INDEX | OCCASIONAL AUTHORS | CITATIONS | H INDEX |
|---|---|---|---|---|---|
| Loet Leydesdorff | **26,484** | 73 | Robert K. Merton | **109,507** | 104 |
| Eugene Garfield | **22,622** | 55 | Francisco Herrera | **38,407** | 101 |
| Mike Thelwall | **13,840** | 61 | Keith Pavitt | **35,521** | 65 |
| Derek J. de Solla Price | **13,263** | 33 | Peter Willett | **25,758** | 74 |
| Francis Narin | **11,297** | 45 | Richard S J Tol | **21,851** | 77 |
| Wolfgang Glänzel | **10,796** | 54 | Stevan Harnad | **17,330** | 62 |
| Ronald Rousseau | **9,570** | 42 | Collins Harry | **16,355** | 49 |
| Chaomei Chen | **9,512** | 43 | Enrique Herrera-Viedma | **16,154** | 62 |
| Anthony F.J. van Raan | **9,200** | 53 | George Kingsley Zipf | **14,745** | 15 |
| Ben R Martin | **8,975** | 39 | Alfred J. Lotka | **14,706** | 30 |
| András Schubert | **8,655** | 45 | Barry Bozeman | **13,764** | 56 |
| Peter Ingwersen | **8,356** | 35 | John Mingers | **11,997** | 49 |
| Henk F. Moed | **8,256** | 46 | Daniele Archibugi | **11,996** | 48 |
| Blaise Cronin | **7,347** | 43 | William C. Clark | **11,915** | 41 |
| Henry Small | **7,307** | 32 | Bart Verspagen | **11,490** | 56 |
| Tibor Braun | **7,231** | 41 | Stan Metcalfe | **10,829** | 50 |
| Vasily V. Nalimov | **6,343** | 31 | Reinhilde Veugelers | **10,581** | 41 |
| Lutz Bornmann | **6,108** | 40 | David I. Stern | **9,695** | 39 |
| Belver C. Griffith | **5,695** | 26 | Yannis Manolopoulos | **9,557** | 45 |
| Howard D. White | **5,569** | 30 | Andy Stirling | **8,989** | 45 |
| Johan Bollen | **5,394** | 33 | Christine L. Borgman | **8,893** | 41 |
| Katy Borner | **5,326** | 31 | Anne-Wil Harzing | **8,839** | 44 |
| Félix de Moya Anegón | **5,074** | 35 | Kal Jarvelin | **8,669** | 32 |
| Koenraad Debackere | **4,933** | 32 | Johan Schot | **8,639** | 32 |
| Jose Maria López Piñero | **4,823** | 31 | John P. Walsh | **8,500** | 29 |

b) Documents

The equivalent list of most influential documents according to GSC in the field of Bibliometrics is available in Table 3.

**Table 3. Top 25 most influential documents in Bibliometrics according to Google Scholar Citations**

| TITLE | AUTHORS | SOURCE | YEAR | CITATIONS |
|---|---|---|---|---|
| Little science, big science | Price | Columbia University Press | 1963 | **5,410** |
| An index to quantify an individual's scientific research output | Hirsch | PNAS | 2005 | **4,860** |
| The dynamics of innovation: from National Systems and "Mode 2" to a Triple Helix of university-industry-government relations | Etzkowitz & Leydesdorff | Research Policy | 2000 | **4,414** |
| Universities and the global knowledge economy: a triple helix of university-industry-government relations | Etzkowitz & Leydesdorff | Pinter Press | 1997 | **2,585** |
| Handbook of Quantitative Science and Technology Research: The Use of Publication and Patent Statistics in Studies of S&T Systems | Moed, Glänzel & Schmoch (ed.) | Springer | 2005 | **2,261** |
| Citation analysis as a tool in journal evaluation. Journals can be ranked by frequency and impact of citations for science policy studies | Garfield | Science | 1972 | **2,166** |
| Citation indexing: Its theory and application in science, technology, and humanities | Garfield | Wiley | 1979 | **2,130** |
| The frequency distribution of scientific productivity | Lotka | J. of Washington Academy Sciences | 1926 | **2,090** |
| Co-citation in the scientific literature: A new measure of the relationship between two documents | Small | JASIS | 1973 | **1,988** |
| Links and impacts: The influence of public research on industrial R&D | Cohen, Nelson & Walsh | Management Science | 2002 | **1,881** |
| Evolution of the social network of scientific collaborations | Barabasi et al | Physica A | 2002 | **1,851** |
| Citation indexes for science. A new dimension in documentation through association of ideas | Garfield | Science | 1955 | **1,783** |
| What is research collaboration? | Katz & Martin | Research Policy | 1997 | **1,591** |
| Handbook of quantitative studies of science and technology | Van Raan (ed.) | North-Holland | 1988 | **1,510** |
| The history and meaning of the journal impact factor | Garfield | JAMA | 2006 | **1,487** |
| The increasing linkage between US technology and public science | Narin, Hamilton & Olivastro | Research Policy | 1997 | **1,211** |
| A general theory of bibliometric and other cumulative advantage processes | Price | JASIST | 1976 | **1,148** |
| Statistical bibliography or bibliometrics? | Pritchard | J. of Documentation | 1969 | **1,134** |
| Theory and practise of the g-index | Egghe | Scientometrics | 2006 | **1,113** |
| The Web of knowledge: a Festschrift in honor of Eugene Garfield | Garfield, Cronin & Atkins (ed). | Information Today | 2000 | **1,102** |
| Visualizing a discipline: An author co-citation analysis of information science, 1972-1995 | White & McCain | JASIS | 1998 | **1,100** |
| CiteSpace II: Detecting and visualizing emerging trends and transient patterns in scientific literature | Chen | JASIST | 2006 | **1,083** |
| Citation analysis in research evaluation | Moed | Springer | 2005 | **1,060** |
| Citation frequency and the value of patented inventions | Harhoff et al | R. of Economics and Statistics | 1999 | **1,023** |
| Maps of random walks on complex networks reveal community structure | Rosvall & Bergstrom | PNAS | 2008 | **992** |

c) Journals

The third unit analysed was the journals in which highly cited documents had been published (i.e., considering only the top 1,000 most cited documents). Table 4 contains the top 25 journals according to the number of highly cited documents published. Additionally, we show the total number of citations received by these articles, the percentage of citations per article (C/A), the percentage of highly cited documents in the sample (HCD) and the distribution of citations.

**Table 4. Top 25 most influential journals in Bibliometrics according to Google Scholar Citations**

| JOURNAL | DOCUMENTS | CITATIONS | C/A | HCD (%) | CITATIONS (%) |
|---|---|---|---|---|---|
| Scientometrics | **284** | 44,384 | 156 | 29.8 | 22.5 |
| JASIST | **137** | 27,021 | 197 | 14.4 | 13.7 |
| Research Policy | **57** | 18,866 | 330 | 6.0 | 9.6 |
| Journal of Informetrics | **36** | 5,052 | 140 | 3.8 | 2.6 |
| Journal of Documentation | **25** | 5,538 | 221 | 2.6 | 2.8 |
| Information Processing & Management | **24** | 4,404 | 183 | 2.5 | 2.2 |
| Journal of Information Science | **20** | 3,815 | 190 | 2.1 | 1.9 |
| Research Evaluation | **18** | 2,126 | 118 | 1.9 | 1.1 |
| ARIST | **14** | 3,621 | 258 | 1.5 | 1.8 |
| Social Studies of Science | **13** | 3,204 | 246 | 1.4 | 1.6 |
| Science and Public Policy | **13** | 2,875 | 221 | 1.4 | 1.5 |
| Plos One | **13** | 2,376 | 182 | 1.4 | 1.2 |
| Nature | **10** | 1,871 | 187 | 1.0 | 1.0 |
| Current Contents | **10** | 1,696 | 169 | 1.0 | 0.9 |
| PNAS | **9** | 7,642 | 849 | 0.9 | 3.9 |
| Science | **8** | 9,219 | 1,152 | 0.8 | 4.7 |
| Library Trends | **7** | 1,230 | 175 | 0.7 | 0.6 |
| Medicina Clinica | **6** | 958 | 159 | 0.6 | 0.5 |
| Online Information Review | **6** | 806 | 134 | 0.6 | 0.4 |
| Science Technology & Human Values | **5** | 946 | 189 | 0.5 | 0.5 |
| Aslib Proceedings | **5** | 765 | 153 | 0.5 | 0.4 |
| Cybermetrics | **5** | 627 | 125 | 0.5 | 0.3 |
| American Psychologist | **4** | 1,026 | 256 | 0,4 | 0,5 |
| World Patent Information | **4** | 726 | 181 | 0.4 | 0.4 |
| Ethics in Science and Environmental Politics | **4** | 687 | 171 | 0.4 | 0.3 |

C/A: Citations per article; HCD (%): Percentage of highly cited articles (top 1,000 most cited documents in the sample; Citations (%): Distribution of citations in the sample

d) Book publishers

The last unit of analysis is the book publishers. The top 20 publishers according to the percentage of highly cited books or book chapters (top 1,000) are presented in Table 5. Additionally, the number of documents, citations (total and percentage of citations respect to the total) and citations per document are displayed.

**Table 5. Top 20 most influential book publishers in Bibliometrics according to Google Scholar Citations**

| PUBLISHER | HCD | HCD (%) | CITATIONS | CITATIONS (%) | C/D |
|---|---|---|---|---|---|
| Springer | 10 | **18,2** | 5,766 | 14,3 | 576.60 |
| Information Today | 6 | **10,9** | 1,635 | 4,0 | 272.50 |
| Wiley | 5 | **9,1** | 3,121 | 7,7 | 624.20 |
| Lexington | 4 | **7,3** | 1,627 | 4,0 | 406.75 |
| Sage | 4 | **7,3** | 1,324 | 3,3 | 331.00 |
| UFMG | 4 | **7,3** | 845 | 2,1 | 211.25 |
| University of Chicago Press | 3 | **5,5** | 6,874 | 17,0 | 2,291.33 |
| Russell Sage Foundation | 3 | **5,5** | 3,836 | 9,5 | 1,278.67 |
| North-Holland | 3 | **5,5** | 2,130 | 5,3 | 710.00 |
| Blackwell | 2 | **3,6** | 1,132 | 2,8 | 566.00 |
| Elsevier | 2 | **3,6** | 1,071 | 2,7 | 535.50 |
| Taylor Graham | 2 | **3,6** | 688 | 1,7 | 344.00 |
| Scarecrow Press | 2 | **3,6** | 416 | 1,0 | 208.00 |
| ISSI | 2 | **3,6** | 276 | 0,7 | 138.00 |
| Ablex | 2 | **3,6** | 193 | 0,5 | 96.50 |
| FECYT | 2 | **3,6** | 193 | 0,5 | 96.50 |
| Columbia University Press | 1 | **1,8** | 5,410 | 13,4 | 5,410.00 |
| Pinter Press | 1 | **1,8** | 2,585 | 6,4 | 2,585.00 |
| Yale University Press | 1 | **1,8** | 936 | 2,3 | 936.00 |
| MIT Press | 1 | **1,8** | 710 | 1,8 | 710.00 |

HCD: Highly cited documents; C/D: Citations per document

## 4.2. The map of the discipline

To visualise the relations between the main actors of Bibliometrics and related fields, a network connecting the main authors and journals/publishers has been generated (Figure 4). Since the set of 1,000 highly cited documents is too big to be easily visualised, only the Top 200 documents have been considered. For each of these documents all authors and sources have been extracted and linked. In this case, all the co-authors of each of the 200 documents have been analysed, discarding authors not related with the discipline, and including authors that are related but do not have a public GSC profile (this approach allows the consideration of this important set of authors, although data from the GS database was needed in addition to the data available in GSC).

**Figure 4. Network of the Bibliometrics discipline through the MADAP method in Google Scholar (author-journal)**

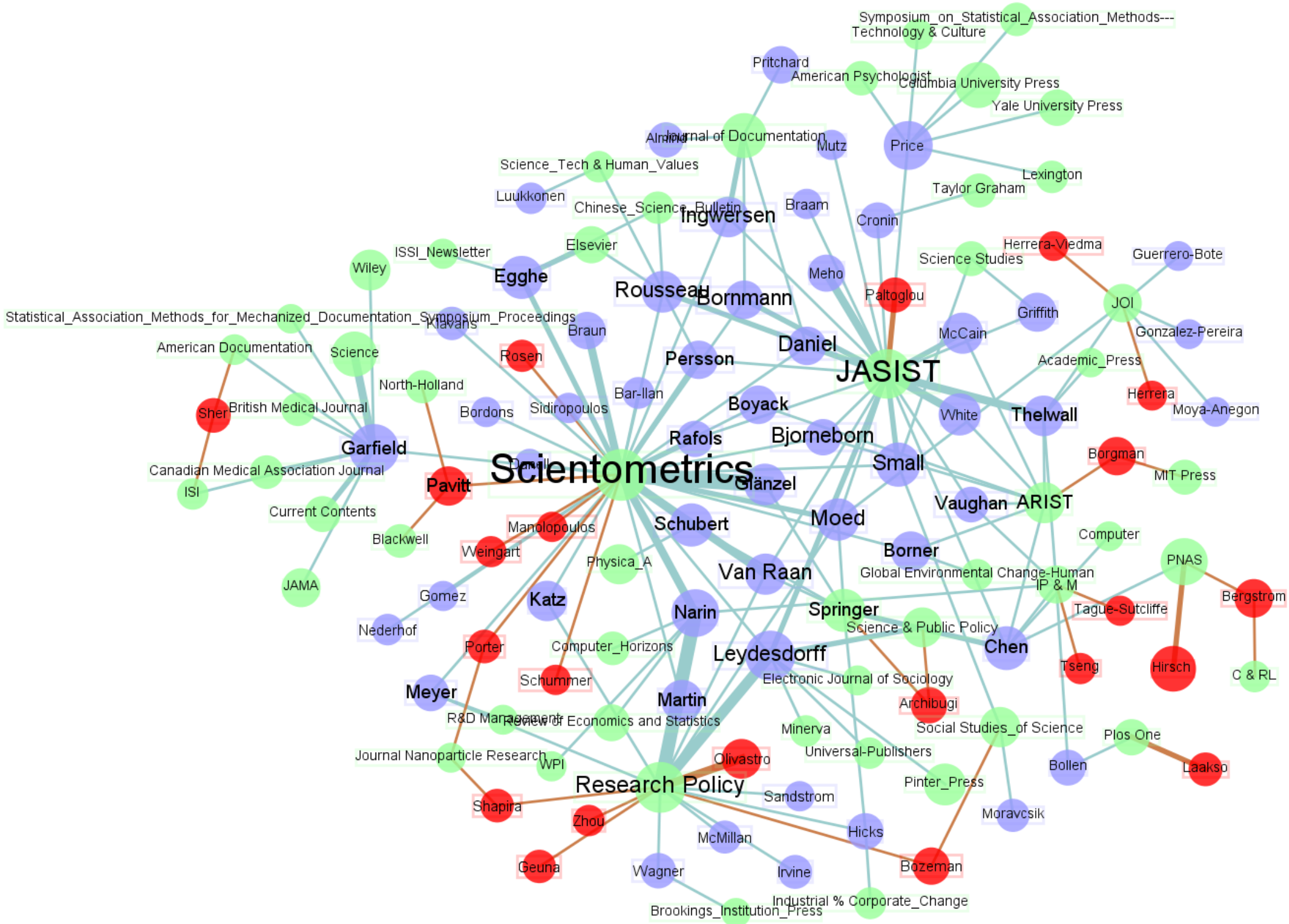

Blue nodes: core authors; Red nodes: related authors; Green nodes: sources
N= 174 nodes (80 sources, 63 core authors, 31 related authors)
Map energysed by Noverlap algorithm with Gephy

The journals with a higher eigenvector centrality are Scientometrics, JASIST and Research Policy. Henk Moed, Loet Leydesdorff, and Anthony Van Raan are the most central specialist authors. Occasional authors (Pavitt, Porter, and Manolopoulos are those with a higher eigenvector centrality score) play a less central role although their influence is notable, especially in relation to some journals (e.g, Research Policy).

Although Figure 4 can reflect author-journal relationships, this map is less informative when it comes to describing sub-disciplines and research fronts. For this reason, an alternative map (Figure 5) has been generated showing author-keyword relationships. In this case, we consider the Top 100 highly cited specialist authors according to GSC public profiles (blue nodes), and all normalized research field keywords included in each of the author profiles (red nodes).

Martín-Martín, A., Orduna-Malea, E., & Delgado López-Cózar, E. (2018). A novel method for depicting academic disciplines through Google Scholar Citations: The case of Bibliometrics. *Scientometrics*, 114(3), 1251–1273. https://doi.org/10.1007/s11192-017-2587-4

**Figure 5. Network of the Bibliometrics discipline through the MADAP method in Google Scholar (author-keyword)**

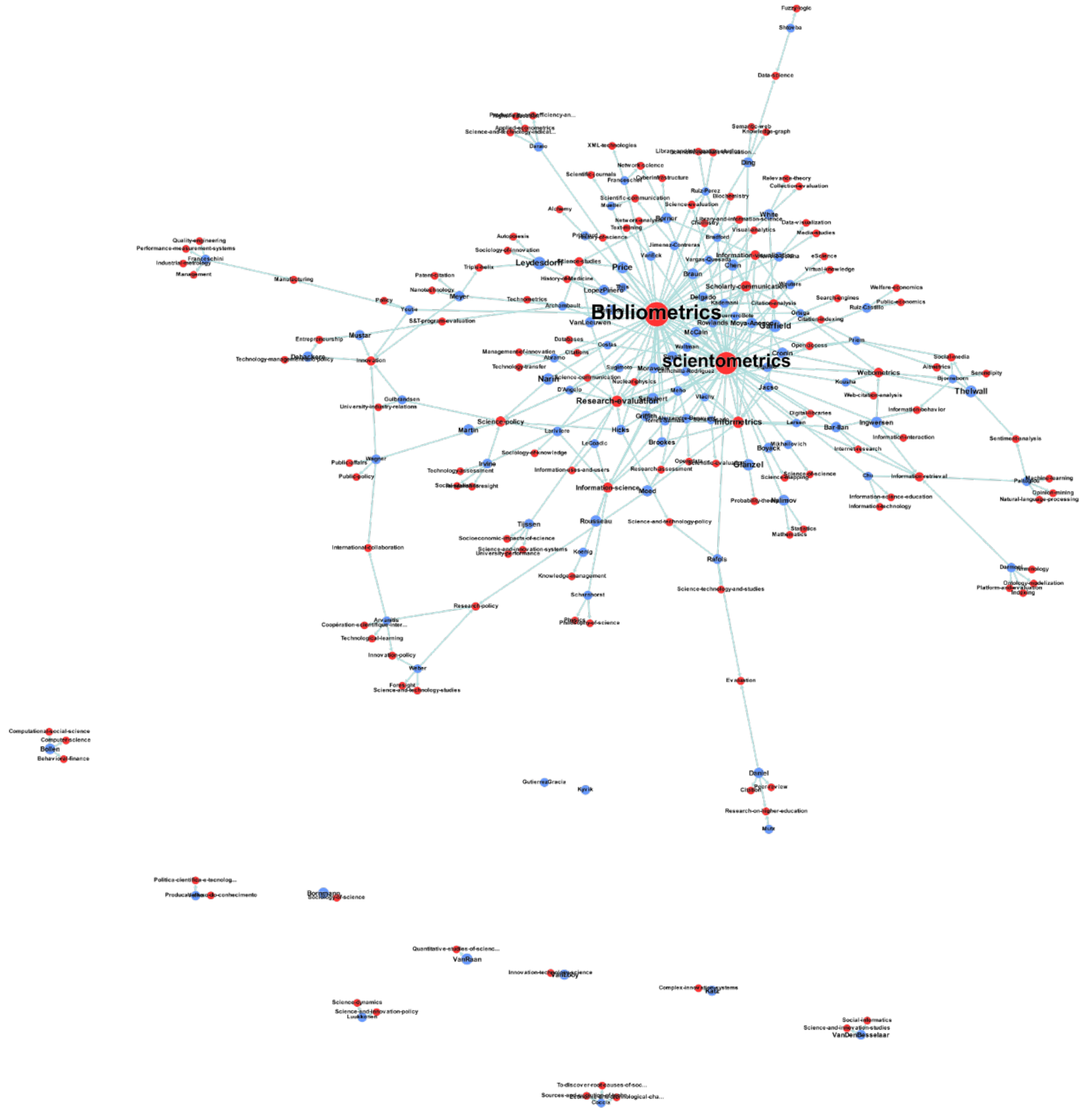

Blue nodes: core authors; Red nodes: topic keywords
N= 239 nodes (100 authors, 139 keywords).
Author node size: times cited; Keyword node size: number of authors sharing the keyword
Map energysed by Force Atlas algorithm with Gephy

This new map groups authors according to the keywords (main research interests) that they selected in their profile, showing sub-disciplinary relationships of the authors (Bibliometrics, Scientometrics, Webometrics, Research evaluation, Science policy, etc.), and at the same time identifying leaders in each front. Additionally, we can observe that some prominent authors with unusual field keywords (e.g., Van Raan or Bornman) are separated from the core, which shows the importance of using appropriate keywords for positioning authors among their peers and creating more accurate disciplinary maps.

## 5. Discussion

### 5.2. About the method (MADAP)

Projects of a bibliographic nature like this one can't ever reach perfection, and it is entirely possible that we may have missed relevant authors. The criteria for selecting the authors were two: first, the existence of a public GSC profile of the author on 24 July 2015 (when the data collection was made), and second, that the author works on the fields of Bibliometrics, Scientometrics, Informetrics, Webometrics, or Altmetrics. Hence, in order to avoid possible confusion, we stress that the ranking of authors (Table 2) was constructed exclusively from the set of 811 authors with a GSC public profile at the time of data collection.

We're well aware that these lists don't include all the researchers in the area. On the one hand some scholars have not created a profile, or they haven't made it public (this is the case of Leo Egghe, an essential figure in the discipline). We should note however that users can create and curate GSC profiles (private preferably) for any researcher, not only for themselves, which may help solving this coverage limitation. Using Harzing's Publish or Perish (PoP) (https://harzing.com/resources/publish-or-perish) in combination with CleanPoP (http://cleanpop.ifris.net) can be an alternative in the cases when a public profile is not available. In addition to this limitation, other scholars may have created a public profile but have included obscure or inadequate keywords to describe their research interests, thus making it impossible to find them using the more common keywords that we used in our approach. We tried to ameliorate this limitation by running the additional topic searches in Google Scholar.

Working with the top cited documents of the discipline – instead of only the authors with a public GSC profile – as the unit of analysis enabled us to capture all relevant authors (whether or not they had a public profile). Documents, journals and publishers rankings (Tables 3, 4, and 5) were constructed following this approach. However, this method requires using Google Scholar in addition to GSC, which adds complexity to the process, is time consuming, and requires a prior in-depth knowledge of the discipline under study. For example, in the case of the network presented in Figure 4 we only analysed the top 200 most cited documents because of these limitations.

Another important point of discussion is the one concerned with the accuracy of data provided by GSC. GSC feeds from GS, which is known to contain errors related both to citation and bibliographic data (recently summarized by Orduna-Malea et al. 2016). These errors are inherited by GSC. However, in GSC authors have the power to edit the bibliographic records and fix these errors. Although it is not likely that many researchers in general bother to do this, the composition of our sample (bibliometricians) makes us think that the data in this particular case might be of a slightly better quality than average. Of course, errors may persist in some profiles. Nevertheless, the manual cleaning process applied in this study prevents bibliographic errors from significantly affecting the general findings.

Another source of errors comes from profile manipulation. Metrics in GSC have been proved to be easily gamed by authors who want to boost their citation counts by abusing self-citations, or by uploading fake academic documents to the Web (Delgado López-Cózar, Robinson-García and Torres-Salinas 2014). Additionally, since GSC profiles can be set to be automatically populated by the system, they may sometimes contain documents that have not been actually authored by the researcher in question (and the researcher may even not be aware of this).

Regarding false citations (caused either by GSC malfunctions or manipulation), their effect in the results obtained in this study is considered to be low, especially on the top positions (the core intellectual map of the discipline). We would like to emphasize that the specific rank positions and metrics in the lists provided (authors, documents, journals, and publishers) should not be considered especially significant. It is the general shape of the discipline that is important. The purpose of this study was to reveal the main agents in the discipline according to the data available in GSC, not to generate micro-level research evaluations.

The main limitation of this method is that it is highly time-consuming. The process of searching, extracting, and cleaning bibliographic data from GS and GSC cannot be completely automated, and much manual labour is required. Carrying out discipline studies with other citation indexes such as Scopus or Web of Science is easier, because they provide more and better metadata. The difference, of course, is that while GS and GSC can be accessed for free, access to Scopus and Web of Science is subject to paying hefty subscription fees. Therefore, each platform presents a tradeoff: with Google Scholar it is possible to freely extract unrefined data. These data requires intensive human intervention to clean in order for it to be useful, which is costly in person-hours. On the other hand, with Scopus and Web of Science it is possible to carry out similar and even more detailed analyses in less time, providing that the necessary (and extremely high) subscription fees have been covered, which is costly in money. The decision of which source is more cost-effective will depend on the type of analyses that need to be carried out, but generally speaking, for small to medium-size projects, the cost of cleaning data extracted from Google Scholar should be several orders of magnitude lower than the subscription costs of the other citation indexes.

Limited time and the availability of just a small workforce are the main reasons why most of this analysis has focused on the most cited documents in the discipline (top 1,000 most cited documents). Thus, this specific analysis mainly presents information on the documents and researchers with the highest impact in the discipline. With more resources (people, time) the analysis could be expanded to cover a larger portion of the data, which would provide insight on the rest of the researchers and their publications. Nevertheless, the method described seems to be a very cost-effective way to accurately represent the structure of the discipline, specially suitable in the cases when accessing other subscription-based citation indexes is not an option.

The extensive coverage in Google Scholar (geographic, linguistic, document types…) is a clear advantage when it comes to developing discipline studies. Particularly, the inclusion of books (see Table 3 and 5) provides a wider vision of the discipline than the one offered by Scopus and Web of Science, where book coverage is merely testimonial (Martin-Martin et al. 2016). In our case, however, 10.5% of the top 200 most highly cited bibliometrics documents according to GS are books (mostly manuals describing techniques and procedures). These documents are not covered by WoS or Scopus.

This method could be used to analyse other disciplines and fields, although as noted before, an in-depth knowledge of the discipline under study may be necessary to identify and contextualize the results obtained. Obviously, the accuracy of the results depends on the level of uptake of the platform by researchers who work in the discipline. It has been reported that coverage of GSC at the discipline level can vary significantly (Ortega, 2015a).

Lastly, the data for this analysis was collected on 2015, and the results would undoubtedly be different if they were collected again now. However, this issue does not compromise the findings of the current study, which were to test the suitability of GSC and GS as sources of data to generate a comprehensive picture of the structure of a discipline, using the procedures previously described (MADAP method).

### 5.2. About the bibliometric actors (the discipline studied)

The accuracy of the method should be discussed not only from a technical/conceptual point of view but also from an empirical perspective. Therefore, we believe it is best to discuss the results obtained from applying the MADAP method to the Bibliometrics field from different points of view (authors, documents, journals, and book publishers).

***Authors***

The top cited authors in Bibliometrics according to GSC (Table 2) accurately represent the map of the discipline, including the founders of the discipline (Price and Garfield) as well as the most influential bibliometricians, almost all of them recipients of the Price medal, a prize that recognizes scientists who have exceptionally contributed with their work to the development of Bibliometrics.

On the one hand, Price, armed with the theoretical foundations laid by John Desmond Bernal and Robert K. Merton, set out to systematically apply quantitative techniques to the History and social studies of Science, developing the theoretical foundations of Scientometrics, born from the combination of the Sociology of science, History, Philosophy of science, and Information science. This approach is characterized by the analysis of the life and activity of Science and scientists from a quantitative perspective. The numbers were used to characterize the production of knowledge and scientists' lives: what they create and produce, to whom they relate to, the sources they used, and the impact and influence they provide/receive to/from other scientists, etc.

On the other hand, Garfield made possible that Bibliometrics became a reality (Bensman 2007; McCain 2010; Small 2017; Wouters 2017): the creation of the "citation index" made possible the quantification of scientific activity through its main output: the publications and citations they generate. Since then, citation analysis and all its variants have become the most widespread analysis technique of this new specialty. This is evidenced by the significant presence of highly cited documents that deal with this topic. Garfield defined the phenotype of the discipline: technology (the basis for the storage and circulation of information) is at the heart of all its tools.

As for the occasional authors of the discipline, these have been included solely as a matter of illustration. Obviously, many of the citations they have received belong to non-bibliometric publications. Nevertheless, the table reflects those important scholars who, despite belonging to other disciplines, provided important contributions to the field. This should be kept in mind when interpreting Table 2.

Lastly, the Bibliometrics map is useful to analyse the rest of the authors in the list: the Hungarian school (both Eastern Europe and Russia, like Nalimov), the Dutch school (with its various branches in Leiden and Amsterdam), the Belgian school (with Egghe and Rousseau), the North American School (Small, Griffith, and White), the Spanish school (with López Piñero, who introduced Price's work in Spain), and the new authors that represent the technological transformation of the discipline (mainly Thelwall).

***Documents***

The top documents in Bibliometrics according to Google Scholar Citations (Table 3) embody the main findings of the field. Among the top documents we can highlight those that first introduced new techniques and citation-based indicators, like the ones by Hirsch (3rd), Garfield (9th and 10th), Small (12th), and Egghe (23rd). Among them we find the most widely known indicator in Bibliometrics (the Impact Factor) and the one that has come to replace it while extending its capabilities (h-index).

The strong orientation of Bibliometrics towards evaluation in general and the assessment of the performance of individuals, journals, and institutions in particular, reveals a clear link between Bibliometrics and Science policy, and explains the use of the aforementioned indicators and other bibliometric tools by policymakers.

Additionally, this list is also a proof of the anomalous institutionalization process of the discipline. The main "bibliometric laws" which still hold true today where established at the dawn of the discipline, even before it was fully instituted (Lotka, Zipf, Bradford), and were developed by authors working outside the discipline. The same happened with the proposal of the h-index by Hirsch, elaborated by this physicist in his "leisure time". Bibliometrics is often revolutionized from outside Bibliometrics.

We can also distinguish the great relevance of some topics such as the "Triple Helix" by Leydersdorff, or the social networks by Barabási, which have had a strong impact outside the borders of our discipline.

Lastly, as we would expect, we can find among the most cited documents those texts that have served as textbooks for the discipline (written by Moed, Van Raan, Eghhe, Rousseau, etc.).

***Journals***

The top journals in Bibliometrics according to GSC (Table 4) illustrate in this case the main communication channels of the discipline.

Scientometrics is the journal with more articles published within the 1,000 most cited documents (284 articles). It is thus the most influential journal in the discipline. Its birth in 1978 was a milestone in the process of institutionalization of the discipline. The second place is occupied by JASIST (137 articles). This fact shows the important role of this journal in Bibliometrics, although its scope is broader. This journal has maintained since its inception a strong link between Information Science and Bibliometrics, though some authors have noticed a slight specialization towards Bibliometrics over time (Nicolaisen and Frandsen 2015). Journal of informetrics, focused exclusively on Bibliometrics, Scientometrics, Webometrics, and Altmetrics, appears in the fourth position (36 articles). The young age of this journal (it was created in 2007) explains why there isn't a greater number of articles published in this journal among the most cited documents in the discipline.

The connection between Library and Information Science (LIS) and Bibliometrics is noticeable through the presence of other important LIS journals in the list, such as Journal of Documentation, Journal of Information Science, Library Trends, or Aslib Proceedings. This connection has been a matter of public record for a long time now (White and McCain 1998; Larivière, Sugimoto and Cronin 2012; Larivière 2012). Its connections with the field of web technologies from an information science perspective is strongly marked as well (Cybermetrics, Online Information Review). Additionally, we can see that journals oriented towards the Social Studies of Science (such as Research Policy, Social Studies of Science, and Science and Public Policy) also have strong ties to Bibliometrics.

If we analyse the number of citations instead of the number of articles published, we find the same first three journals occupying the first positions (Scientometrics, JASIST, and Research Policy), but the data also shows a great impact of articles published outside the core journals of the discipline, revealing the role of multidisciplinary journals. Science gets 9,219 citations from only 8 articles whereas PNAS gets 7,642 citations from 9 articles, and PLoS One gets 2,376 citations from 13 articles (the figures for Nature are lower, with 1,871 citations from 10 articles).

As regards the contributions published outside both the core and multidisciplinary journals (primarily bibliometric studies of specific fields published in the journals of the field), the MADAP method is able to capture both the documents and journals only if at least one of the co-authors of these manuscripts have been previously identified by the search and identification process (See section 3.1), and have created a GSC public profile. In this sense, the method does not exclude these contributions by default.

***Book publishers***

In this case, output is low (the first position is occupied by Springer, with only 10 documents positioned within the set of highly cited documents), although we observe that all publishers achieve high numbers of citations per document (Springer receives 5,766 citations to 10 documents). Also remarkable is the performance of university presses in the dissemination of bibliometric research results (such as the University of Chicago, Columbia, Yale or MIT), with a very low presence in terms of productivity but an impressive impact in the number of citations. The ability to attract well-established authors in order to publish specialized books makes a great difference in book publisher rankings.

Martín-Martín, A., Orduna-Malea, E., & Delgado López-Cózar, E. (2018). A novel method for depicting academic disciplines through Google Scholar Citations: The case of Bibliometrics. *Scientometrics*, 114(3), 1251–1273. https://doi.org/10.1007/s11192-017-2587-4

## 6. Conclusions

By virtue of the results obtained, the research question (RQ1) can be answered positively. GSC (in combination with Google Scholar) is able to provide a precise and accurate picture of the Bibliometrics community. Moreover, the data collected, not only at the author-level but also at the document-level and source-level, clearly responds to our mental image of the field. That is, it is possible to identify the most influential authors (both specialists and occasional researchers), documents (articles and books) and sources (journals and publishers) in the discipline using data from GSC. Therefore, the MADAP method has been proved not only feasible but also accurate and valid (RQ2).

The application of the procedures followed in this work (the MADAP method) to study other fields and disciplines through GSC challenges new research on this front.

## Acknowledgements

Funding was provided by Ministerio de Educación, Cultura y Deporte (FPU2013/05863), Universitat Politècnica de València (PAID-10-14).